\documentclass[12pt]{article}
\usepackage{pic04}
\usepackage{hyperref}
\usepackage{url}
\usepackage{graphicx}

\begin{document}

\title{\bf HIGH ENERGY GAMMA RAYS}
\author{
Mathieu de Naurois       \\
{\em LPNHE IN2P3 - CNRS - Universities Paris VI-VII,}\\
{\em 4 place Jussieu,  75252 Paris Cedex 05 FRANCE}}
\maketitle

%
%
%
%
%
%
\vspace{4.5cm}
%

\baselineskip=14.5pt
\begin{abstract}
The Very High Energy Gamma Ray Astronomy (VHE) is a rapidly evolving 
branch of modern astronomy, which covers the range from about 50 GeV
to several tens of TeV from the ground. 
In the past years, the second generation instruments firmly established 
a growing and varied list of  sources  including plerions, supernova
remnants and active galactic nuclei, and started to study some fundamental
questions such as the origin of cosmic rays or the emission mechanisms
of the active galactic nuclei.

New results now include the first VHE unidentified sources as well as 
more puzzling sources such as the Galactic center. The arrival of new 
generation instruments (HESS, CANGAROO III, VERITAS, MAGIC) already
gives a impressive look at the near future.
Here we attempt to summarize the current status of the field. We 
briefly describe the instruments and analysis techniques, and
give an outlook on the sources detected sofar.
\end{abstract}
\newpage

\baselineskip=17pt

\section{The universe in gamma-rays}

\subsection{A little bit of history}

Since their discovery by Victor Hess in 1912, one of the most puzzling (and still not 
completely solved) problems of astronomy is the {\it origin of Cosmic Rays}.
These high energy particles mostly consist of charged nuclei and are spread over 
more than 10 orders of magnitude in energy up to $10^{20}\,\mathrm{eV}$ (and possibly above).
The search for the corresponding cosmic accelerators motivated the developement 
of {\it gamma-ray astronomy}, first from space and then from the ground. 
In contrast to the charged component of the cosmic rays, the gamma-rays are not
deviated by the Galactic and extra-galactic magnetic fields and thus point back
to their emission source. They are emitted by particle physics processes 
(non-thermal synchrotron radiation of accelerated electrons, inverse Compton scattering 
of cosmic rays off ambient photons, pion decay,\dots) occuring in 
high electromagnetic field acceleration regions, shocks in astrophysical
plasmas or interaction of the cosmic rays with the interstellar medium. 
In the active galactic nuclei or other compact objects, they are 
thought to be emitted by the very central engine (instead of being secondary 
products) and can therefore probably provide the most valuable constraints on 
the emission models. Finally, they could also possibly originate from annihilation 
of neutralinos in dark matter clumps.

\begin{figure}[htb]
\begin{center}
\includegraphics[width=7cm,angle=90]{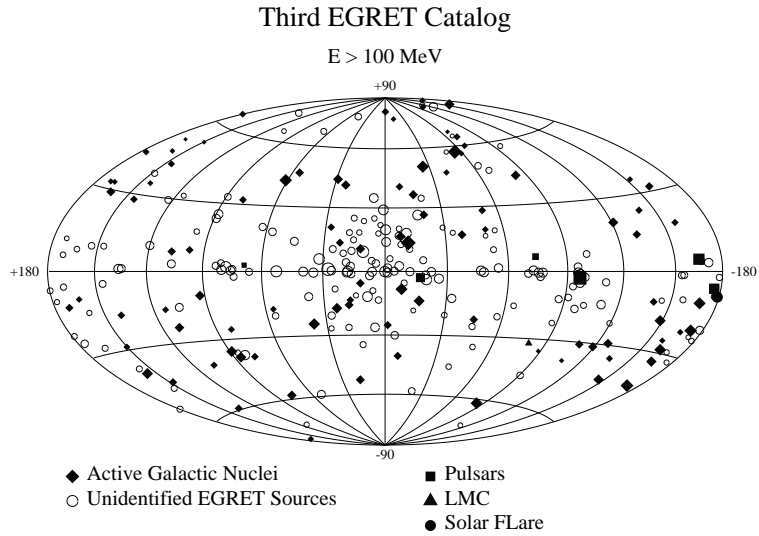}
\end{center}
\vspace{-2em}
\caption{\it Third Egret Catalog. \label{EGRET}}
\end{figure}

After the pionneering work of SAS II and COS-B, a breakthrough was achieved 
in the 1990's by the EGRET detector onboard the CGRO satellite: this 
spark-chamber detector made the first map of the diffuse
$\gamma$-ray emission from the Galactic plane and detected 271 point-like source
between 100 MeV and 10 GeV. The third EGRET catalogue \cite{EgretCat} (Figure \ref{EGRET})
contains 66 AGNs, 8 pulsars and 170 still-unidentified sources. A important fraction
of these sources should be Galactic, and probably contribute to the bulk of 
cosmic-ray sources. The identification of these sources is one of the major
chalenges of the $21^\mathrm{th}$ century astronomy.

\subsection{Atmospheric Cerenkov Imaging and other techniques}

\begin{figure}[htb]
\begin{center}
\centerline{\hbox{ \hspace{0.2cm}
\includegraphics[width=6.5cm]{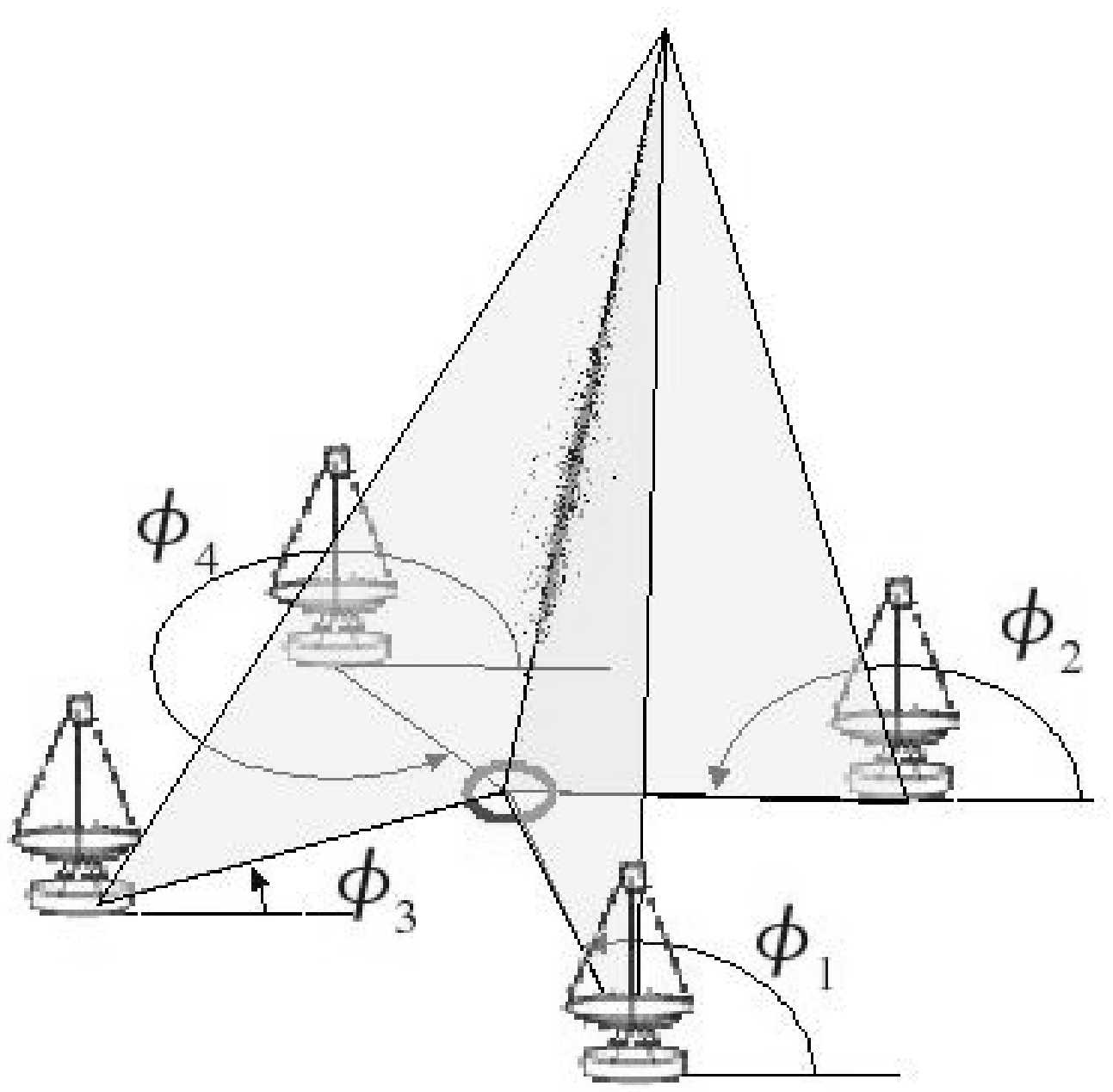}
\hspace{0.3cm}
\includegraphics[width=6.5cm]{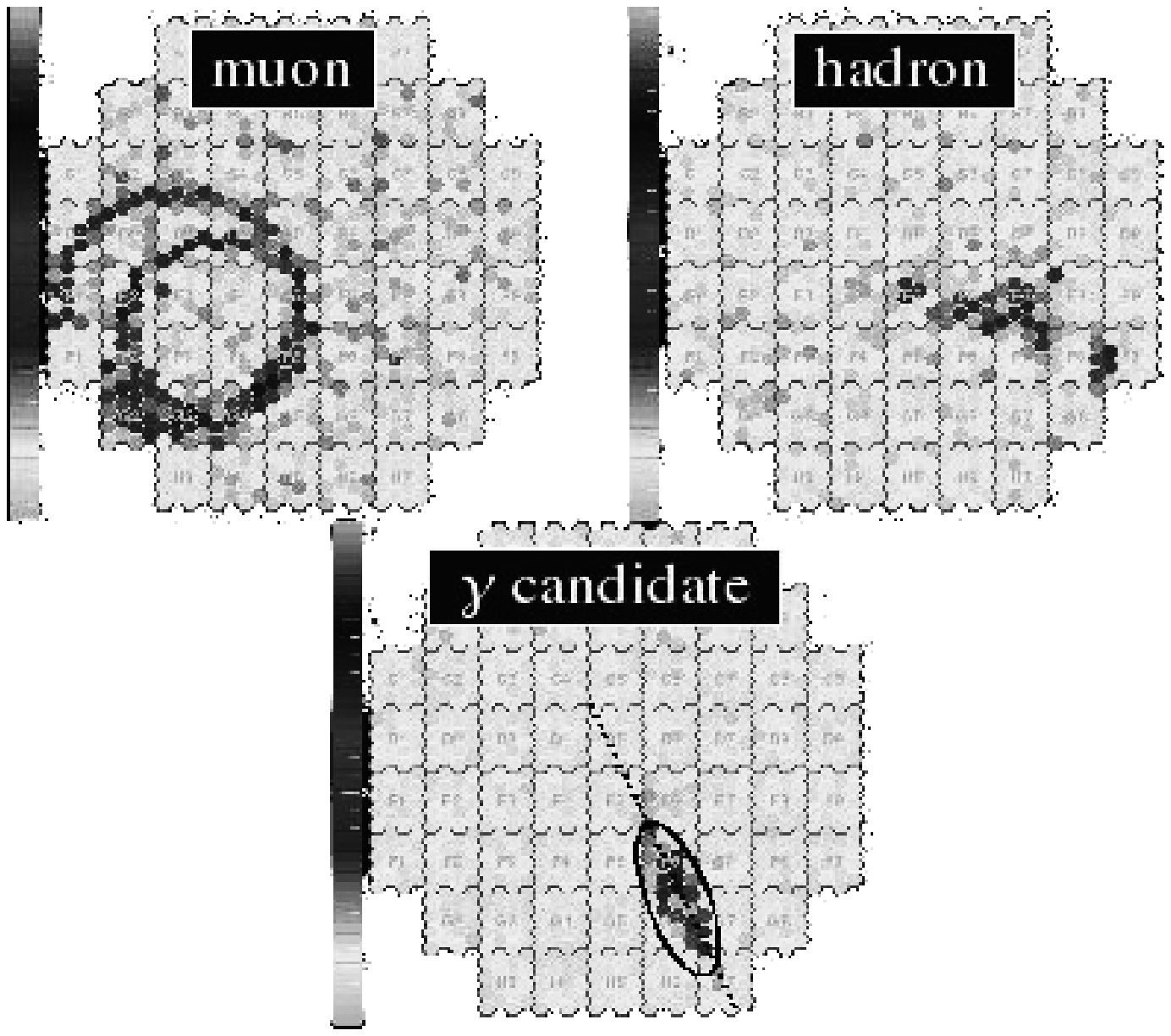}
} }
\caption{\it The Atmospheric Cerenkov Imaging technique. Left:
detection principle. Right: Typical images for muons, hadron and
$\gamma$ candidates. \label{IACT}}
\vspace{-2em}
\end{center}
\end{figure}

Above 10 GeV the rapid fall in of the flux limits the sensitivity of
space detectors.
The ground base detection relies on the sampling of  the Cerenkov light emitted 
by the charged particles in the extensive air showers (Figure \ref{IACT}). 
The opening angle of the Cerenkov light makes the showers visible at great 
distances (up to 300 m) away from the detector and thus permits huge detection 
area ($>10^9\ \mathrm{cm}^2$). The shape of the image is then used 
to discriminate between the high cosmic ray background (several hundreds Hz) 
and the $\gamma$ candidates (representing at most $0.2\%$ of the
background for the strongest sources).

Between roughly $20\ \mathrm{GeV}$ and $100\ \mathrm{GeV}$ remains a
quasi-unexplored region, where the Cerenkov light is too faint for existing
Imaging Cerenkov Telescopes. Several Solar Farm experiments (Solar II, Stacee,
Celeste) took the opportunity of the huge light collection area provided
by the existing solar plants to lower the threshold down to $50\ \mathrm{GeV}$, 
but they could never reach high-enough hadron rejection capabilities to be competitive.

\begin{figure}[htb]
\begin{center}
\includegraphics[height=5cm]{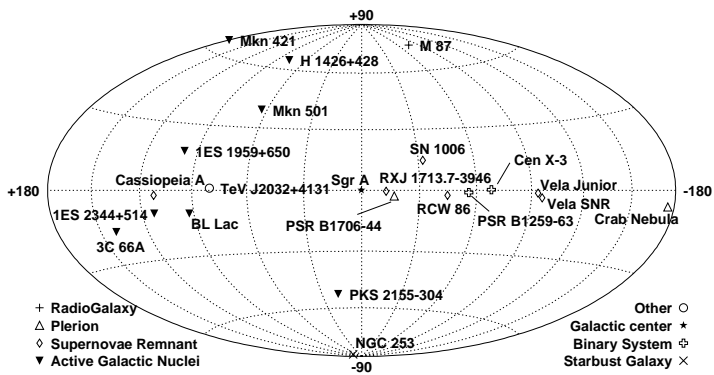}
\end{center}
\vspace{-1em}
\caption{\it The VHE sky c. 2004 \label{VHE}}
\vspace{-2em}
\end{figure}

The VHE catalogue (figure \ref{VHE}) was opened in 1989 by
the discovery of the Crab Nebula \cite{WHIPPLECrab} (see section \ref{sec:Crab}) by the WHIPPLE collaboration.
HEGRA introduced in 1995 the stereoscopic technique, which, by looking at the same shower
from different points of view, gave a dramatic improvement in hadron and local muon
rejection as well as in angular and energy resolution.
After 15 years the VHE catalogue now consists of about 20 sources with very varied 
properties.

\subsection{New instruments}

\begin{table}[htb]
\vspace{-1.5em}
\centering
\caption{ \it Third generation instruments.\label{ThirdGeneration}}
\vskip 0.1 in
\begin{tabular}{|l|c|c|c|c|c|} \hline
 Name        &  Location & \# Telescopes & Diameter & FoV & Opening \\
\hline
\hline
 CANGAROO III & Australia & 4 & 10~m & $4^\circ$ & Completed\\
 HESS   & Namibia & 4 & 12~m & $5^\circ$ & Completed\\
 MAGIC & La Palma & 1 & 17~m & $4^\circ$ & Commissionning \\
 VERITAS & USA & 4$\rightarrow$7 & 12~m & $3.5^\circ$ & 2006 \\
\hline
\end{tabular}
\label{extab}
\vspace{-0.5em}
\end{table}

After the success of the previous generation (WHIPPLE, HEGRA, CAT, CANGAROO-II),
third generation instruments are now coming online. Their key properties are summarized
in the table \ref{ThirdGeneration}. Three of these projects are multi-telescope arrays
combining the advantages of a big reflector inherited from WHIPPLE, 
a fine pixelization camera first developed by CAT, and stereoscopic observation
pionneered by HEGRA. Their detection threshold lies around 100 GeV at zenith, and increases
with zenith angle due to bigger atmospheric thickness.
The latest experiment, MAGIC, is a bigger dish single telescope 
experiment which comprises a number of advanced technologies in the design of the 
mirror and the signal transmission, and aims to lower the detection threshold
closer to the 20 GeV domain. 
Complementary to this, survey instrument with poor point-like sensitivity
but nearly full sky coverage and duty cycle (Milagro, Tibet III) are providing 
the firsts VHE surveys at higher energies (above $2\ \mathrm{TeV}$).

\section{Galactic source}

Nearly half of the known VHE sources belongs to our Galaxy. We will briefly
recall the properties of some of them, focusing on the most recent results.

\subsection{The Crab Nebula\label{sec:Crab}}

\begin{figure}[htb]
\begin{center}
\centerline{\hbox{ \hspace{0.2cm}
\includegraphics[width=5cm]{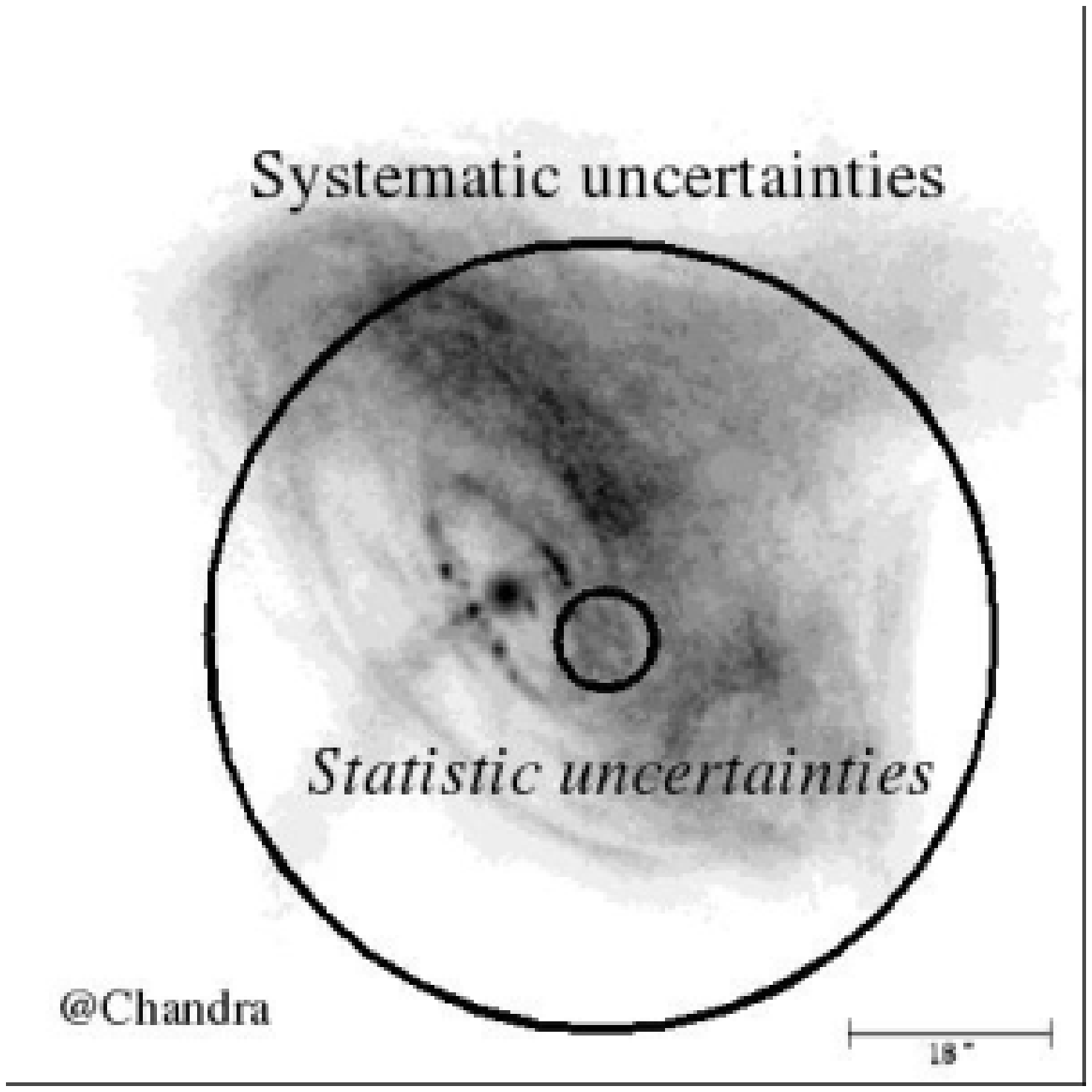}
\hspace{0.3cm}
\includegraphics[width=7.5cm]{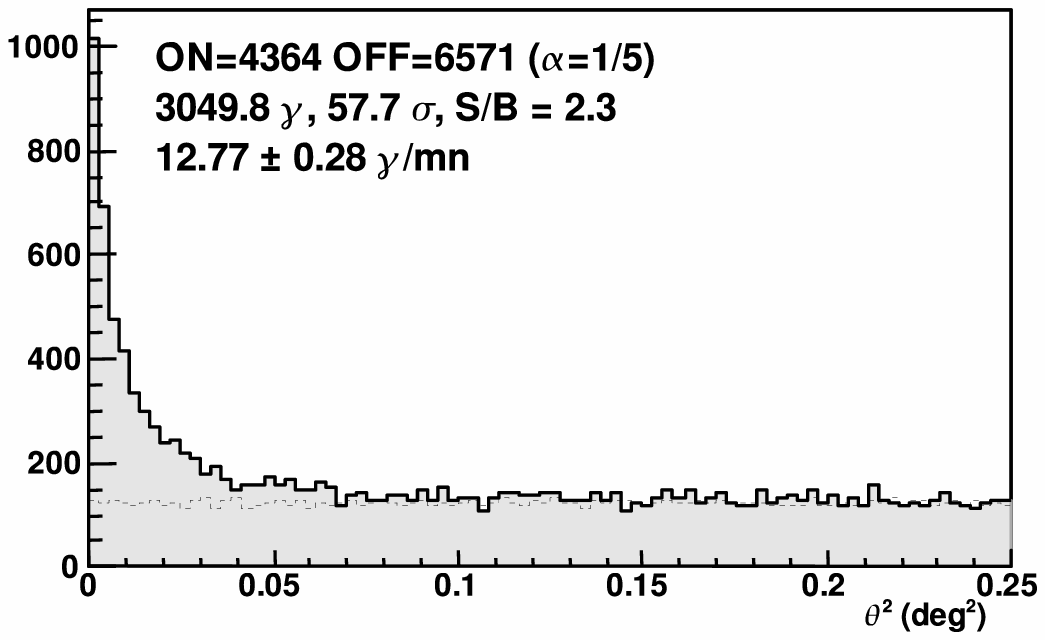}
} }
\caption{\it The Crab Nebula. Left: high resolution X-rays image from Chandra,
with confidence region from HESS. Right: Distribution of squared angular 
distance of the events to the direction of the Crab nebula as recorded by HESS, showing a 
highly significant excess in the direction of the source ($\theta^2 = 0$). \label{CrabResult}}
\vspace{-2.5em}
\end{center}
\end{figure}

This first discovered VHE source is a {\it plerion}, that is
a synchrotron nebula fed by the electron wind of a central pulsar.
A high resolution image of the Crab Nebula by the telescope Chandra is show 
in figure \ref{CrabResult}. Since its first discovery in 1989, the Crab Nebula
showed no evidence for variability of any kind. It is therefore considered
as the {\it standard candle} of high energy gamma-ray astronomy and can be used
to compare and intercalibrate the instruments.

New observation carried out by HESS using an uncomplete array of 3 telescopes
yielded a highly significant signal, as show in the figure \ref{CrabResult}.
The supreme achieved angular resolution gives a position compatible with the
central pulsar, however a definite conclusion on the emission region will 
be possible only in a near future, when the systematic uncertainties will be reduced. 
The spectrum derived by HESS is found to be compatible with previously published results. MAGIC also reported
a $10~\sigma$ detection of the Crab Nebula during the comissionning phase.

\subsection{Galactic Pevatrons: the Supernova remnants}

The supernova remnants are though to be the site of acceleration of the 
Galactic cosmic rays up to $10^{15}\,\mathrm{eV}$. However the naive picture
according to which the TeV spectrum would give a clear signature of
$\pi^\circ$ decay has not turned true, and the situation is still under debate.

\begin{figure}[htb]
\begin{center}
\centerline{\hbox{ \hspace{0.2cm}
\includegraphics[width=4cm]{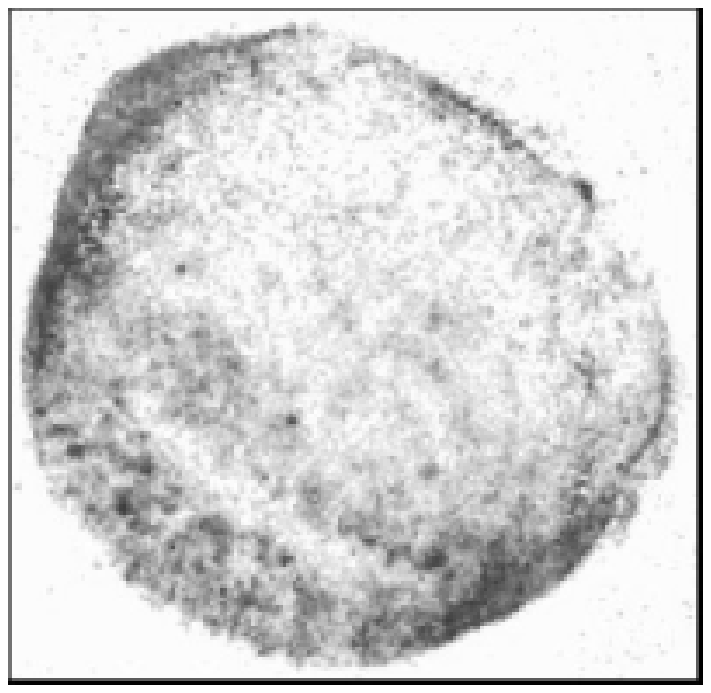}
\hspace{0.3cm}
\includegraphics[width=4cm]{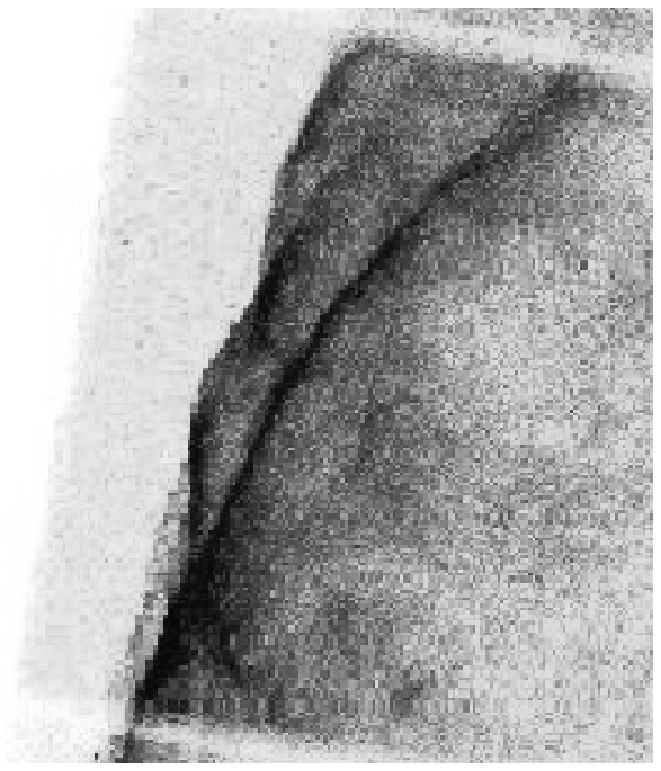}
\hspace{0.3cm}
\includegraphics[width=5.5cm]{Chandra_ShockThickness.eps}
} }
\caption{\it Left: The SN 1006 supernova remnant seen by Rosat. Middle: zoom on the north west region by Chandra \cite{BambaSN1006}. 
Right: section of one of the filaments showing the shock thinckness. \label{ChandraSN1006}}
\vspace{-2.5em}
\end{center}
\end{figure}

The detection of the supernova remnant SN~1006 by CANGAROO-II in 1996 and 1997 was originally
attributed to an inverse Compton emission of accelerated electrons, without the need
of nuclear cosmic rays. The situation changed in 2003 with the obervation by Chandra \cite{BambaSN1006}
of very thin filaments of intense non thermal synchrotron X-ray emission. This has been interpreted
as the effect of magnetic field amplification made possible by a accelerated population of 
nuclear cosmic rays \cite{BerezhkoSN1006}. In this scheme of high magnetic field ($\geq 100 \mu G$),
the TeV emission would be mainly due to $\pi^\circ$ decay. 

HESS reported a non detection of SN~1006 and derived an upper limit at the level 
of 10\% of the published CANGAROO-II flux. This discrepancy raises a lot of questions: 
As the emission of a SNR is not expected to vary significantly on so short timescales,
could the emission come rather from a extragalactic background source? Only new observations
will solved this point.

CANGAROO-II recently reported the detection of two other supernova remnants, RCW~86 and RX~J0852.0-4622. 
The supernova remnants are now firmly established 
as sources of $100\ \mathrm{TeV}$ electrons, but the debate concerning the nuclear cosmic rays 
acceleration is not closed yet. We expect the situation to improve very quickly 
with the results of the third generation instruments.

\subsection{An exotic system: PSR B1259-63}

\begin{figure}[htb]
\begin{center}
\centerline{\hbox{ \hspace{0.2cm}
\includegraphics[width=7cm]{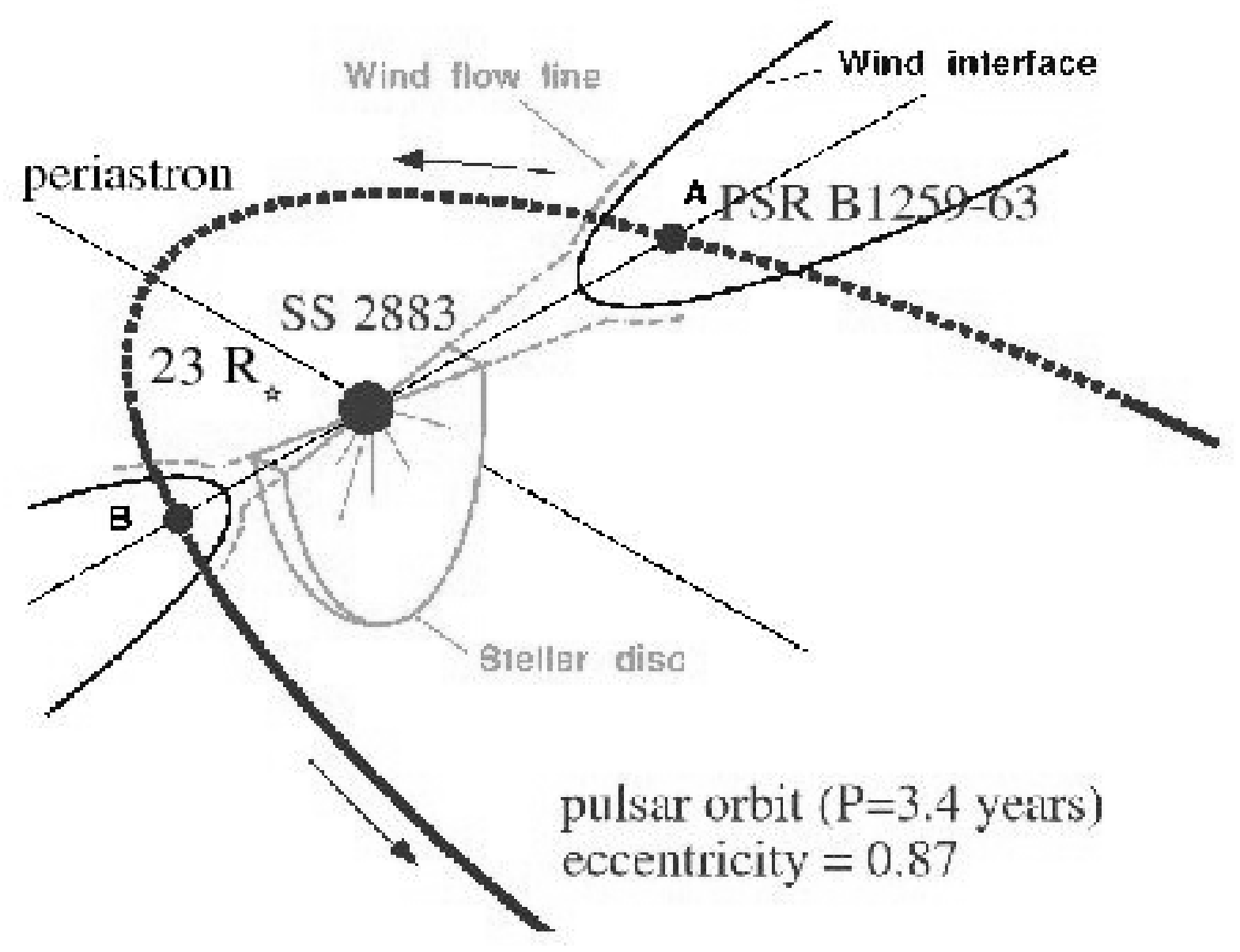}
\hspace{0.3cm}
\includegraphics[width=7.5cm]{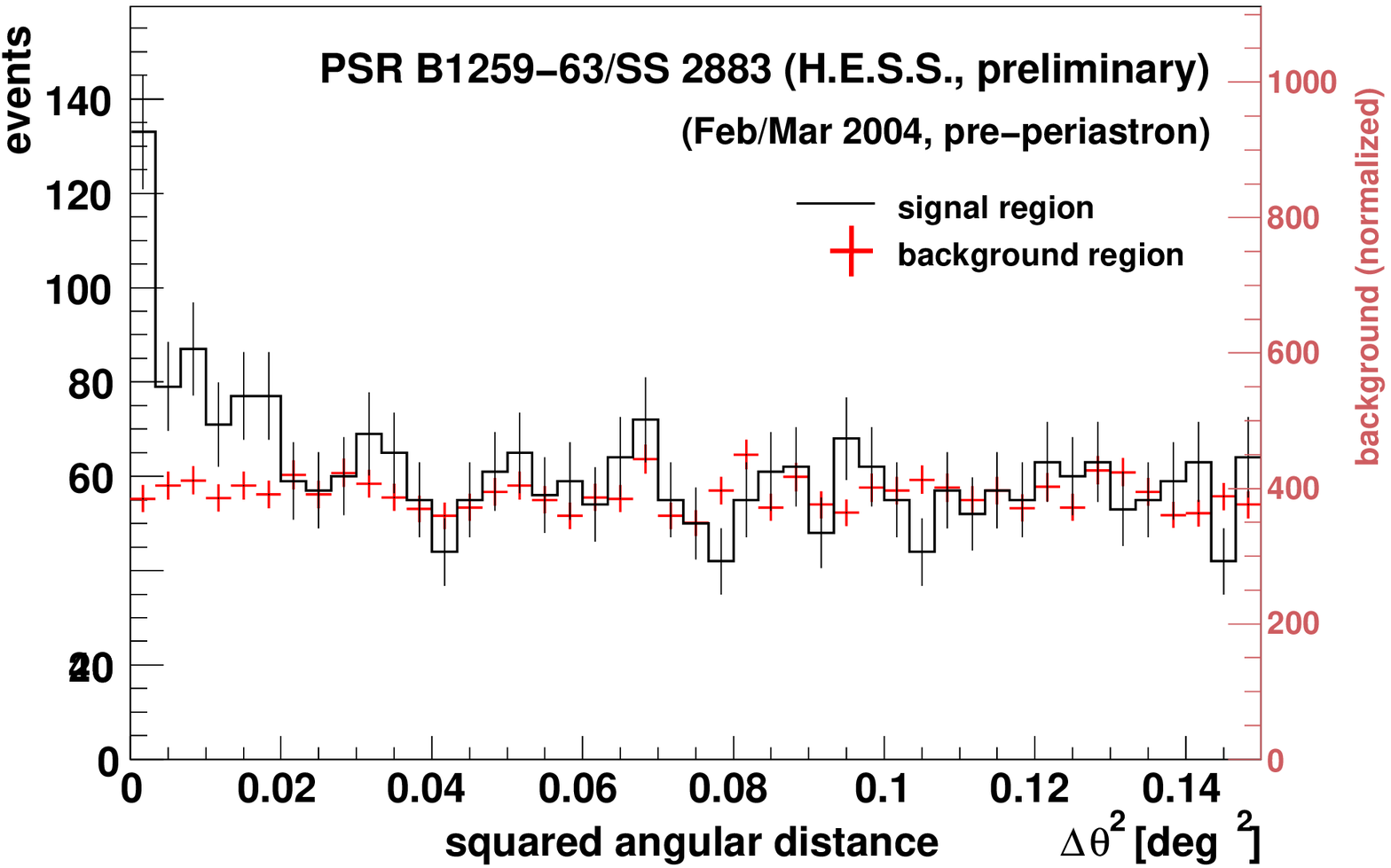}
} }
\caption{\it Left: schematics of the PSR B1259-63 binary system. Right: signal observed by HESS\label{PSRB1259}}
\vspace{-2em}
\end{center}
\end{figure}

PSR B1259-63 is a very unique system in our Galaxy, consisting of a $47.7\ \mathrm{ms}$ 
pulsar in highly ($\epsilon \approx 0.87$) excentric orbit around a massive 
star SS~2883 (figure \ref{PSRB1259}). Every 3.4 years at the {\it periastron}, 
the distance of the pulsar to the star is only $~ 23R_\star$, $R_\star$ being
the radius of the star. An interaction between the pulsar wind and the stellar disk
surrounding SS~2883 could, according to several models (e.g. \cite{Kirk}), lead 
to the production of high energy gamma-rays.

HESS observed PSR B1259-63 during 10 hours around its periastron (March, 7th 2004).
An significant excess at the $8.2~\sigma$ has been detected, and further work on the 
modeling is under way.

\subsection{The Galactic center}

The CANGAROO-II experiment first announced the discovery of a TeV gamma-ray signal from
the Galactic center \cite{SgrACANGAROO}. Soon after, WHIPPLE brought a marginal 
confirmation \cite{SgaWHIPPLE}. The very soft and somewhat unusual spectrum measured
by CANGAROO-II, $\propto E^{-4.6}$ and the compatibility of the position with the center
of the Galaxy led many authors to interpret this signal in terms of neutralinos 
annihilation \cite{Hooper}. For a very massive WIMP (of mass $m_\chi$) annihilating mostly into 
$W^+W^-$, $Z^\circ Z^\circ$ and $q\bar q$ pairs, a reasonable description of the decay spectrum
is given by the parametrisation of the form:

\begin{equation}
\frac{dN}{dE} = \frac{0.73}{m_\chi}\times \frac{e^{-7.8 E/m_\chi}}{E_0 + (E/m_\chi)^1.5}
\label{eq:Wimp}
\end{equation}

If the signal observed by CANGAROO-II was due to annihilation of neutralinos, the mass of the
later would be in the range $1-3\ \mathrm{TeV}$.

\begin{figure}[htb]
\begin{center}
\includegraphics[width=7cm]{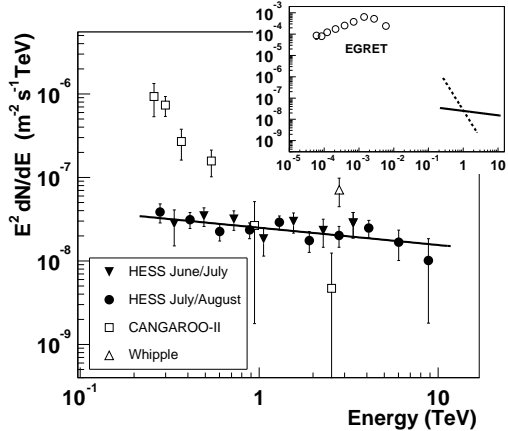}
\end{center}
\vspace{-2em}
\caption{\it CANGAROO-II and HESS spectra of the Galactic Center. \label{SgrAFig}}
\end{figure}

As shown in the figure \ref{SgrAFig}, recent results obtained by 
HESS \cite{HESSSagA} with a much better angular resolution (better than $0.1^\circ$) 
are incompatible with CANGAROO-II, and would give, if the signal was due to neutralinos, a lower 
limit of $7.5\ \mathrm{TeV}$ on the neutralino mass at $99\%~\mathrm{CL}$. However, 
the Galactic center is a very crowded region, with several potential sources in the 
CANGAROO-II and even in the HESS confidence level region. 
New observations with the complete HESS array should for instance clearly exclude the 
supernova remnant SgrA East which lies only
$1.7'$ away from the Galactic center.

More conventional model for the emission of central black hole $\mathrm{Sgr~A}^\star$ include Advection
Dominated Accretion Flow (ADAF), or diffuse emission due to interaction of accelerated
protons and nuclei with the high density ambient matter.

\subsection{TeV~2032+42 : First TeV unidentifed source}

During the year 2002 the first unidentified 
TeV source, {\bf TeV~2032+42} was detected by HEGRA \cite{TeV2032}. 
This extended and faint source (3\% of the Crab flux), having no radio or X-rays counterparts, exhibits a 
hard spectrum and has been confirmed by WHIPPLE analysis of archival data
at a slighlty higher level.  
There are now some indications that this is not an isolated case, 
but rather a hint of a bright future.

\section{A quick look outside the Galaxy}

Almost all known extragalactic TeV sources are active galactic nuclei (AGN) belonging
to the class of {\bf Blazars}. They consist of a super-massive $(\approx 10^9\,M_\odot)$ black
holes surrounded by an accretion disk, from which two giant ultrarelativistic jets
of plasma escape up to Megaparsec distances. For the blazars, one of theses jets 
points towards the earth and completely outshines the rest of the AGN.

Amongst other properties, the TeV emission of the Blazars is characterized by a dramatic
variability on all time scales, the flux been sometimes multiplied by a factor 2 in times
as short as 20 minutes.

Two principle classes of models aim to explain the emission mechanism of the
Blazars: The leptonic SSC ({\it Synchroton Self Compton}) and EC ({\it Extrernal Compton}) 
models attribute the TeV emission to inverse Compton emission of accelerated electrons
respectively on the synchroton X photons or on the environment photons. The second
class of models is based on proton acceleration and hadronic cascades in the jets.

Very detailed observations of the correlation and timelag between the X-ray and $\gamma$-ray 
emissions are required to discriminate between the models. Recently organized multi-wavelength
observation campains, combining the radio, optical, X-rays and $\gamma$-rays observations,
tend to give a more comprehensive picture in favour of the leptonic models.

Very recent observations of giant flares of Markarian 421 by HESS and MAGIC 
confirm that the new generation instruments now reached the required sensitivity
to study the time-variability of the Blazars at the minute timescale, and will
probably solve this acceleration problem in a short time-scale. 

\section{Conclusion}

\begin{figure}[htb]
\begin{center}
\vspace{-2em}
\includegraphics[width=13cm]{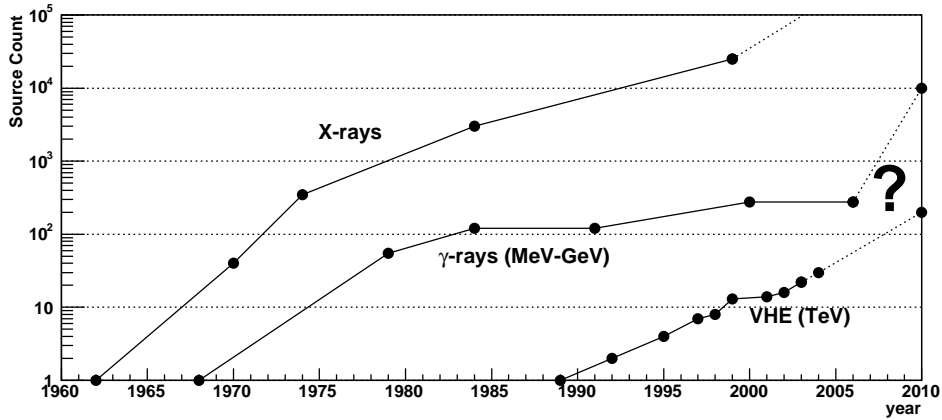}
\end{center}
\vspace{-2em}
\caption{\it Moore's law of astonomy: Source count vs time for X-ray, $\gamma$-ray
and high energy $\gamma$-ray astronomy. \label{SourceCount}}
\end{figure}

In the last years, the VHE astronomy left the field of experiment to become a real branch
of astonomy. The VHE catalogue already contains a big variety of sources and steadily 
increases. The similar evolution of the source count versus time for X-ray, $\gamma$-ray 
and high energy $\gamma$-ray astronomy (Figure \ref{SourceCount}) gives hints about a very 
bright future. The increase of statistics about extragalactic sources will also
make new measurements possible: The interaction of the $\gamma$-rays  with intergalactic 
infra-red and visible photons (through electron-positron pair creation) can be
exploited to do a tomography of the ambiant star light at cosmological
distances. This very delicate measurement would require a significance number of 
AGN with similar spectral features to disentangle the intrinsic properties
of each source from the effect of comic absorption.

\section{Acknowledgements}
I would like to thank the organizers of the conference for their
very warm welcome.

\end{document}